\documentclass[twocolumn,amsmath,nofootinbib,superscriptaddress]{revtex4} 

\usepackage{url}
\usepackage{amssymb,aas_macros}
\usepackage{amsfonts}
\usepackage{amsbsy}
\usepackage{graphicx}
\usepackage{hyperref}
\usepackage{array} 
\usepackage{multirow}

\newcolumntype{x}[1]{%
>{\centering\hspace{0pt}}p{#1}}%
\providecommand{\openone}{\leavevmode\hbox{\small1\kern-3.8pt\normalsize1}}

\def\ie{{\frenchspacing\it i.e.}}
\def\eg{{\frenchspacing\it e.g.}}

\def\spose#1{\hbox to 0pt{#1\hss}}
\def\simlt{\mathrel{\spose{\lower 3pt\hbox{$\mathchar"218$}}
   \raise 2.0pt\hbox{$\mathchar"13C$}}}
\def\simgt{\mathrel{\spose{\lower 3pt\hbox{$\mathchar"218$}}
     \raise 2.0pt\hbox{$\mathchar"13E$}}}
 \def\simpropto{\mathrel{\spose{\lower 3pt\hbox{$\mathchar"218$}}
     \raise 2.0pt\hbox{$\propto$}}}

\def\beq#1{\begin{equation}\label{#1}}
\def\eeq{\end{equation}}
\def\beqa#1{\begin{eqnarray}\label{#1}}
\def\eeqa{\end{eqnarray}}

\def\fig#1{Figure~\ref{#1}}

\def\ed{\end{document}}

\def\rn{}
\def\nn#1 #2{#2. #1}				
\def\nnn#1 #2 #3{#2. #3. #1}			
\def\nnnn#1 #2 #3 #4{#2. #3. #4 #1}		
\def\nnnnn#1 #2 #3 #4 #5{#2. #3. #4 #5. #1}	

\def\rf#1;#2;#3;#4;#5 {{\frenchspacing\par\rn#1, #3 {\bf #4}, #5 (#2). \par}}
\def\rg#1;#2;#3;#4;#5;#6 {{\frenchspacing\par\rn#1, #3 {\bf #4}, #5 (#2). \par}}
\def\rfbook#1;#2;#3;#4;#5 {{\frenchspacing\par\rn#1, {\it #3} (#5, #4, #2).\par}}
\def\rfprep#1;#2;#3 {{\par\frenchspacing\rn#1, #3 (#2).\par}}
\def\rfproc#1;#2;#3;#4;#5;#6 {{\frenchspacing\par\rn#1 #2, in {\it #3}, ed. #4 (#5: #6)\par}}
\def\rfprocp#1;#2;#3;#4;#5;#6;#7 {{\frenchspacing\par\rn#1 #2, in {\it #3}, ed. #4 (#5: #6), p#7\par}}

\begin{document}
\pdfoptionalwaysusepdfpagebox=5


\title{Mapping our Universe in 3D with MITEoR\footnote{To be published in proceedings of 2013 IEEE International Symposium on Phased Array Systems \& Technology}}

\author{Haoxuan Zheng}
\affiliation{Dept.~of Physics \& MIT Kavli Institute, Massachusetts Institute of Technology, Cambridge, MA 02139}
\email{jeff\_z@mit.edu}
\author{Max Tegmark}
\affiliation{Dept.~of Physics \& MIT Kavli Institute, Massachusetts Institute of Technology, Cambridge, MA 02139}

\author{Victor Buza}
\affiliation{Dept.~of Physics \& MIT Kavli Institute, Massachusetts Institute of Technology, Cambridge, MA 02139}
\affiliation{Dept.~of Physics, Harvard University, Cambridge, MA 02138}

\author{Joshua S.~Dillon}
\affiliation{Dept.~of Physics \& MIT Kavli Institute, Massachusetts Institute of Technology, Cambridge, MA 02139}

\author{Hrant Gharibyan}
\affiliation{Dept.~of Physics \& MIT Kavli Institute, Massachusetts Institute of Technology, Cambridge, MA 02139}

\author{Jack Hickish}
\affiliation{Dept.~of Physics, University of Oxford, Oxford, OX1 3RH, United Kingdom}

\author{Eben Kunz}
\affiliation{Dept.~of Physics \& MIT Kavli Institute, Massachusetts Institute of Technology, Cambridge, MA 02139}

\author{Adrian Liu}
\affiliation{Dept.~of Physics \& MIT Kavli Institute, Massachusetts Institute of Technology, Cambridge, MA 02139}
\affiliation{Dept.~of Astronomy and Radio Astronomy Lab, University of California, Berkeley, CA, 94720}

\author{Jon Losh}
\affiliation{Dept.~of Physics \& MIT Kavli Institute, Massachusetts Institute of Technology, Cambridge, MA 02139}

\author{Andrew Lutomirski}
\affiliation{Dept.~of Physics \& MIT Kavli Institute, Massachusetts Institute of Technology, Cambridge, MA 02139}

\author{Scott Morrison}
\affiliation{Dept.~of Physics \& MIT Kavli Institute, Massachusetts Institute of Technology, Cambridge, MA 02139}

\author{Sruthi Narayanan}
\affiliation{Dept.~of Physics \& MIT Kavli Institute, Massachusetts Institute of Technology, Cambridge, MA 02139}

\author{Ashley Perko}
\affiliation{Dept.~of Physics \& MIT Kavli Institute, Massachusetts Institute of Technology, Cambridge, MA 02139}
\affiliation{Dept.~of Physics, Stanford University, Stanford, CA, 94305}

\author{Devon Rosner}
\affiliation{Dept.~of Physics \& MIT Kavli Institute, Massachusetts Institute of Technology, Cambridge, MA 02139}

\author{Nevada Sanchez}
\affiliation{Dept.~of Physics \& MIT Kavli Institute, Massachusetts Institute of Technology, Cambridge, MA 02139}

\author{Katelin Schutz}
\affiliation{Dept.~of Physics \& MIT Kavli Institute, Massachusetts Institute of Technology, Cambridge, MA 02139}

\author{Shana M.~Tribiano}
\affiliation{Dept.~of Physics \& MIT Kavli Institute, Massachusetts Institute of Technology, Cambridge, MA 02139}
\affiliation{Science Dept.~Borough of Manhattan Community College, City University of New York, New York, 10007}

\author{Matias Zaldarriaga}
\affiliation{School of Natural Sciences, Institute for Advanced Study, Princeton, New Jersey, 08540}

\author{Kristian Zarb Adami}
\affiliation{Dept.~of Physics, University of Oxford, Oxford, OX1 3RH, United Kingdom}

\author{Ioana Zelko}
\affiliation{Dept.~of Physics \& MIT Kavli Institute, Massachusetts Institute of Technology, Cambridge, MA 02139}

\author{Kevin Zheng}
\affiliation{Dept.~of Physics \& MIT Kavli Institute, Massachusetts Institute of Technology, Cambridge, MA 02139}

\author{Richard Armstrong}
\affiliation{Dept.~of Physics, University of Oxford, Oxford, OX1 3RH, United Kingdom}

\author{Richard F.~Bradley}
\affiliation{Dept.~of Electrical and Computer Engineering, University of Virginia, Charlottesville, VA, 22904}
\affiliation{National Radio Astronomy Observatory, Charlottesville, VA, 22903}

\author{Matthew R.~Dexter}
\affiliation{Dept.~of Astronomy and Radio Astronomy Lab, University of California, Berkeley, CA, 94720}

\author{Aaron Ewall-Wice}
\affiliation{Dept.~of Physics \& MIT Kavli Institute, Massachusetts Institute of Technology, Cambridge, MA 02139}

\author{Alessio Magro}
\affiliation{Dept.~of Physics, University of Malta, Msida MSD 2080, Malta}

\author{Michael Matejek}
\affiliation{Dept.~of Physics \& MIT Kavli Institute, Massachusetts Institute of Technology, Cambridge, MA 02139}

\author{Edward Morgan}
\affiliation{Dept.~of Physics \& MIT Kavli Institute, Massachusetts Institute of Technology, Cambridge, MA 02139}

\author{Abraham R. Neben}
\affiliation{Dept.~of Physics \& MIT Kavli Institute, Massachusetts Institute of Technology, Cambridge, MA 02139}

\author{Qinxuan Pan}
\affiliation{Dept.~of Physics \& MIT Kavli Institute, Massachusetts Institute of Technology, Cambridge, MA 02139}

\author{Courtney M.~Peterson}
\affiliation{Dept.~of Physics, Harvard University, Cambridge, MA 02138}

\author{Meng Su}
\affiliation{Dept.~of Physics \& MIT Kavli Institute, Massachusetts Institute of Technology, Cambridge, MA 02139}

\author{Joel Villasenor}
\affiliation{Dept.~of Physics \& MIT Kavli Institute, Massachusetts Institute of Technology, Cambridge, MA 02139}

\author{Christopher L.~Williams}
\affiliation{Dept.~of Physics \& MIT Kavli Institute, Massachusetts Institute of Technology, Cambridge, MA 02139}

\author{Hung-I Yang}
\affiliation{Dept.~of Physics \& MIT Kavli Institute, Massachusetts Institute of Technology, Cambridge, MA 02139}

\author{Yan Zhu}
\affiliation{Dept.~of Physics \& MIT Kavli Institute, Massachusetts Institute of Technology, Cambridge, MA 02139}
\affiliation{Dept.~of Physics, Stanford University, Stanford, CA, 94305}

\date{\today}

\vspace{10mm}

\begin{abstract}
Mapping our universe in 3D by imaging the redshifted 21 cm line from neutral hydrogen has the potential to overtake the cosmic microwave background as our most powerful cosmological probe, because it can map a much larger volume of our Universe, shedding new light on the epoch of reionization, inflation, dark matter, dark energy, and neutrino masses. 
We report on MITEoR, a pathfinder low-frequency radio interferometer whose goal is to test technologies that greatly reduce the cost of such 3D mapping for a given sensitivity.  MITEoR accomplishes this by using massive baseline redundancy both to enable automated precision calibration and to cut the correlator cost scaling from $N^2$ to 
$N\log N$, where $N$ is the number of antennas.
The success of MITEoR with its 64 dual-polarization elements bodes well for the more ambitious HERA project, which would incorporate many identical or similar technologies using an order of magnitude more antennas, each with dramatically larger collecting area.
\end{abstract} 

\maketitle

\section{Introduction}

\begin{figure}[!ht]
\centerline{\includegraphics[width=80mm]{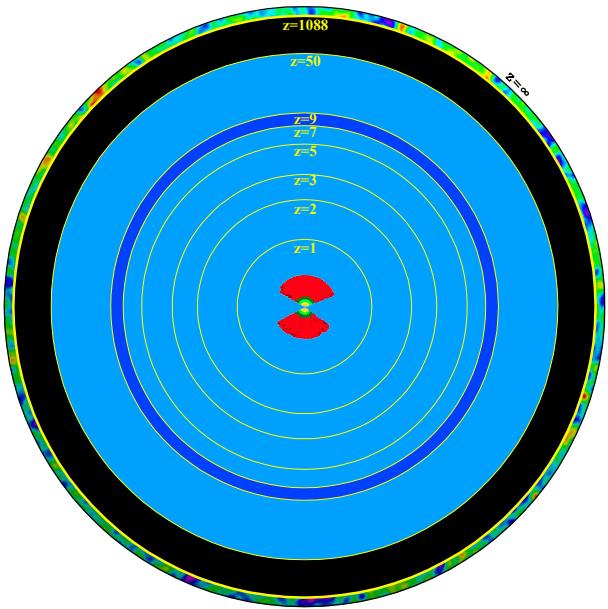}}
\caption{
21 cm tomography can potentially map most of our observable universe (light blue/grey), whereas the CMB probes mainly a thin 
shell at $z\approx 1100$ and current large-scale structure maps (here exemplified by the Sloan Digital Sky Survey 
and its luminous red galaxies) map only small volumes near the center. Half of the comoving volume lies at $z>29$.
Even the convenient $7\simlt z \simlt 9$ region (dark blue/grey) can eclipse the CMB in cosmological precision \cite{Yi}, 
probing the nature of neutrinos, dark energy, dark matter, reionization and early universe.}
\label{HubbleVolumeFig}
\end{figure}

Mapping neutral hydrogen throughout our Universe via its redshifted 21 cm line 
offers a unique opportunity to probe the cosmic ``dark ages'' and the formation of the first luminous objects.
Moreover, because it can map a much larger volume of our Universe (\fig{HubbleVolumeFig}), 
it has the potential to overtake the 
Cosmic Microwave Background (CMB) as our most sensitive cosmological probe of the epoch of reionization, inflation, dark matter, dark energy, and neutrino masses. For example, a square kilometer radio array with maximal sky coverage could improve the sensitivity to spatial curvature and neutrino masses by up
to two orders of magnitude, to $\Delta\Omega_k\approx 0.0002$ and $\Delta m_\nu\approx 0.007$ eV, and shed new light on the early universe by a $4\sigma$ detection of the spectral index running predicted by the simplest  inflation models \cite{Yi}.

A first challenge is that the cosmological 21cm signal is so faint that none of the currently competing experiments around the world (LOFAR \cite{LOFAR}, MWA \cite{MWA}, PAPER \cite{PAPER}, 21CMA \cite{21cma}, GMRT \cite{GMRT})
have yet detected it, although increasingly stringent upper limits have recently been placed \cite{GMRT2,MWAJosh,PAPERpspec}. A second challenge is that foreground contamination from our galaxy and extragalactic sources is perhaps four orders of magnitude larger than the cosmological hydrogen signal \cite{GSM} and must be accurately cleaned out from the data, which requires even greater sensitivity as well more accurate calibration and beam modeling than the current state-of-the-art in radio astronomy (see \cite{FurlanettoReview,miguelreview} for reviews).

Large sensitivity requires large collecting area, which unfortunately requires a large budget. Since steerable single dish radio telescopes become prohibitively expensive beyond a certain point, the above-mentioned current experiments have all opted for interferometry, combining a large number $N$ of independent antenna elements which are (except for GMRT) individually more affordable. For example, the MWA experiment has built  $N=128$ ``tiles'', each consisting of 16 dipole-like antennas that cost less than \$100 apiece---including the low-noise amplifiers. The LOFAR, MWA, PAPER, 21CMA and GMRT experiments currently have comparable $N$.
The challenge is that  all of these experiments use standard hardware cross-correlators whose cost grows quadratically with $N$, since they need to correlate all $N(N-1)/2\sim N^2/2$ pairs of antenna elements. This cost is reasonable for the current scale $N\sim 10^2$, but will completely dominate the cost for $N\simgt 10^3$, making precision cosmology arrays with $N\sim 10^6$ as discussed in \cite{Yi} completely unfeasible.

\begin{figure}[pbt]
\vskip-1.6cm
\centerline{\includegraphics[width=103mm]{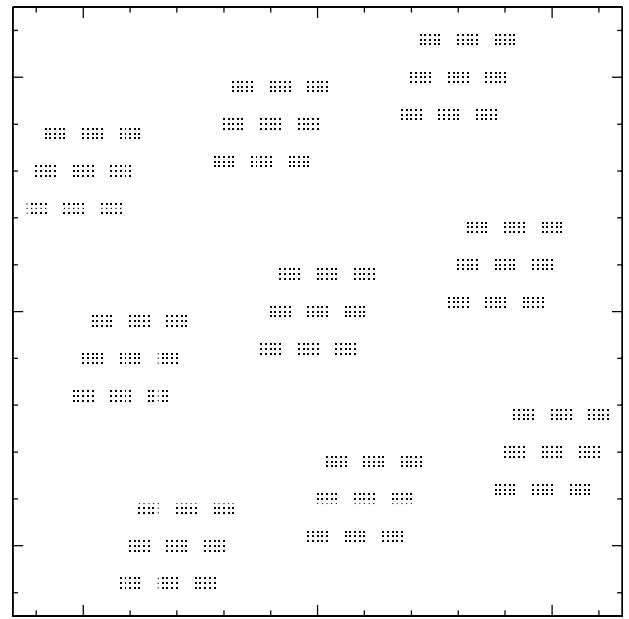}}
\vskip-3.3cm\caption{
A hierarchical grid omniscope correlator allows much more general antenna layouts as well. The figure shows an example of a 3-level hierarchy of antennas, with $5\times 3$ blocks arranged in $3\times 3$ blocks that are in turn 
placed in a $3\times 3$ block. Note that the blocks at each level can be non-square (like at level 1), sheared (like at level 2) and rotated (like at level 3), in any combination. It was shown in \cite{FFTT2} that this particular array can be efficiently correlated with a 6-dimensional FFT.
\label{LayoutExample}
}
\end{figure}

For the particular application of 21 cm cosmology, however, designs with better cost scaling are possible, as we described in \cite{FFTT}: by arranging the antennas in a 
compact rectangular grid and performing the correlations using Fast Fourier Transforms (FFTs), thereby cutting the cost scaling to $N\log N$.\footnote{This basic idea traces to the early days of radio astronomy .
Initially the Fourier transforms were done by analog means and usually in only one dimension 
(\eg, using a so-called Butler matrix \cite{Butler61}), severely limiting the number of antennas that could be used.
For example, the 45 MHz interferometer in \cite{May84} used six elements. 
A Japanese group worked on an analog $8\times 8$ FFT Telescope about 15 years ago for studying transient radio sources
\cite{Nakajima92,Nakajima93}, and then upgraded it to digital signal processing aiming for a $16\times 16$ array with a field of view just under  $1^\circ$. 
Electronics from this effort is also used in the 1-dimensional 8-element Nasu Interferometer \cite{Takeuchi05}.
}
Moreover, some of us recently showed \cite{FFTT2} that the class of antenna layouts allowing cheap correlators (with cost scaling like $N\log N$ correlators) is encouragingly large, including not only previously discussed rectangular grids but also arbitrary hierarchies of such grids, 
with arbitrary rotations and shears at each level (as in \fig{LayoutExample}): all correlations for such a 2D array with an $n$-level hierarchy 
can be efficiently computed via a Fast Fourier Transform in not $2$ but $2n$ dimensions.

This is particularly attractive for science applications requiring exquisite sensitivity at vastly different angular scales, such as
21cm cosmology (where short baselines are needed to probe the cosmological signal\footnote{It has been shown that the 21cm signal-to-noise (S/N) per resolution element in the $uv$-plane (Fourier plane) is $\ll 1$ for all current 21cm cosmology experiments, and that their cosmological sensitivity therefore improves by moving their antennas closer together to focus on the center of the $uv$-plane and bringing its S/N closer to unity
\cite{Matt3,Yi}. Error bars on the cosmological power spectrum have contributions from both noise and sample variance, and it is well-known that the total error bars on a given physical scale (for a fixed experimental cost) are minimized when both contributions are comparable, which happens when the $S/N\sim 1$ on that scale. This is why more compact 21cm experiments have been advocated.
This is also why early suborbital CMB experiments focused on small patches of sky to get S/N$\sim 1$ per pixel, and why galaxy redshift surveys target objects like luminous red galaxies that give S/N$\sim 1$ per 3D voxel.
} 
and long baselines are needed for point source removal).
Such hierarchical grids thus combine the angular resolution advantage of traditional array layouts with the cost advantage of a rectangular Fast Fourier Transform Telescope.
Since individual antennas are digitized separately rather than grouped into tiles,  $N$ independent sky beams are imaged simultaneously, thus improving all-sky sensitivity by imaging almost half the sky at any one time as long as the antenna elements have very broad primary beams. In contrast, the MWA and 21CMA tiles form only one beam at a time, and the LOFAR tiles initially form only four.
If the antennas have a broad spectral response as well and their signals are digitized with high bandwidth, the cosmological neutral hydrogen gets simultaneously imaged in a vast 3D volume covering both much of the sky and also a vast range of distances (corresponding to different redshifts, \ie, different observed frequencies.)
We will refer to such a low-cost array that is effectively omnidirectional and omnichromatic as an {\bf omniscope}.

A second advantage of such hierarchical grid arrays is that they have massive numbers of redundant baselines, which can be used to improve calibration.
Building on past redundant baseline calibration methods by Wieringa \cite{Wieringa} and others, we recently developed an algorithm which is both automatic and statistically unbiased, able to produce precision phase and gain calibration for all antennas in a hierarchical grid without making any assumptions about the sky signal \cite{omnical}.  Similar algorithms have been implemented and tested in \cite{LOFARcal,PAPERpspec}.  Once obtained, precision calibration solutions can in turn produce more accurate modeling of the synthesized and primary beams\footnote{For tile-based interferometers like the MWA and 21CMA, gain and phase errors in individual antennas do not get regularly calibrated, adding a fundamental uncertainty to the tile sky response.} \cite{JonniePrimaryBeam}, which has been shown to improve the quality of the foreground modeling and removal which is so crucial to 21 cm cosmology.

\section{The MITEoR Experiment}

Niels Bohr allegedly said: {\it ``In theory, theory and practice are the same. In practice, they're not."}
It is therefore timely to develop a pathfinder instrument that tests the omniscope approach to 3D hydrogen mapping and assesses how well it works in practice. The MIT Epoch-of-Reionization experiment (MITEoR) is such a pathfinder instrument, designed to test this approach and assess how well it works in practice. 
The goal of this paper is to describe MITEoR and provide a brief progress report. 

\subsection{System overview}

\begin{figure}[pbt]
\centerline{\includegraphics[width=85mm]{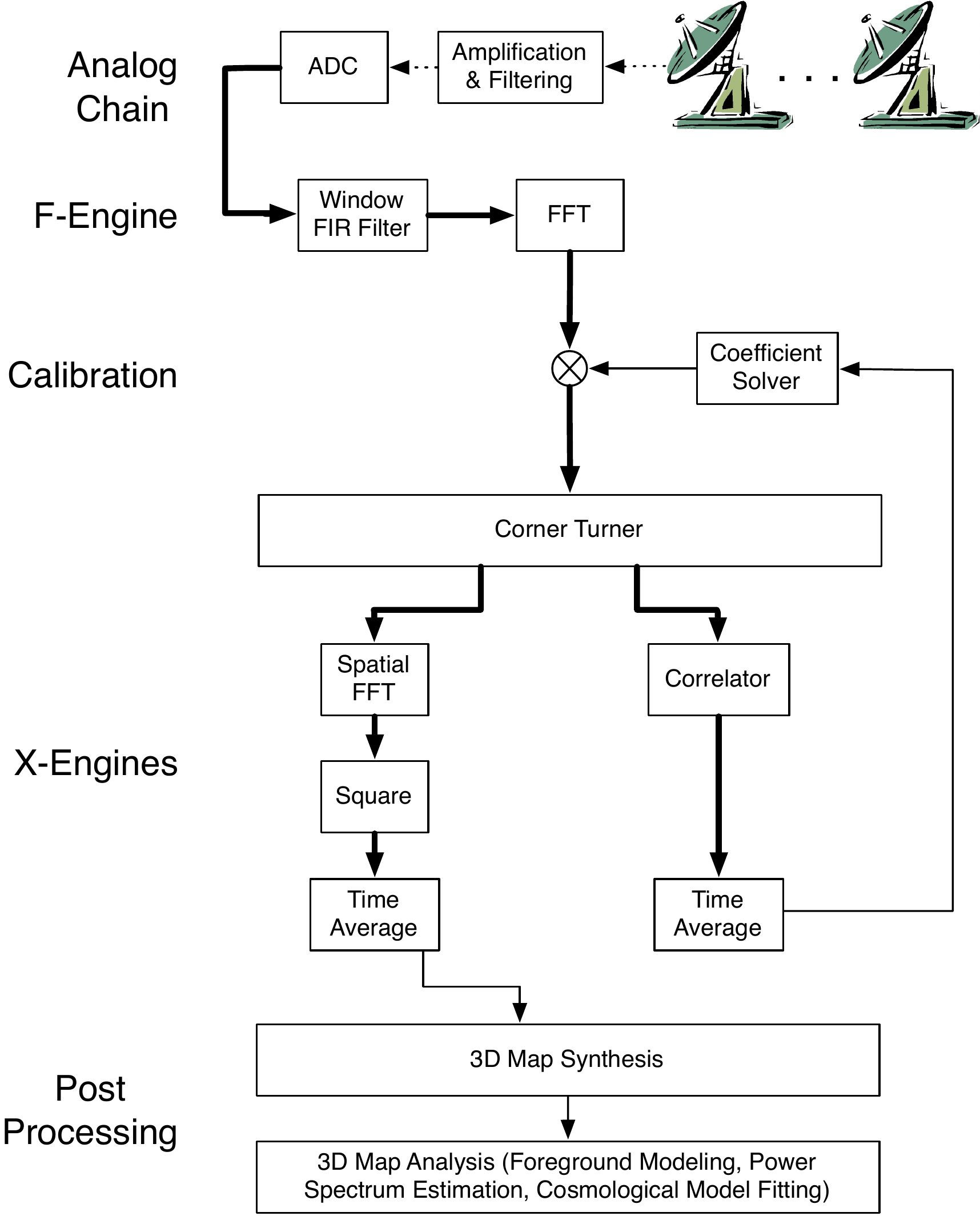}}
\caption{
Data flow in a very large omniscope.
A hierarchical grid of dual-polarization antennas converts the sky signal into volts, which get amplified and filtered by the analog chain, transported to a central location, and converted into bits every few nanoseconds.
These high-volume digital signals (thick lines) get processed by field-programmable gate arrays (FPGAs) which perform a temporal Fourier transform (using polyphase filter banks). The FPGAs (or GPUs) then multiply by complex-valued calibration coefficients that depend on antenna, polarization and frequency, then spatially Fourier transform, square and accumulate the results, recording integrated sky snapshots every few seconds and thus reducing the data rate by a factor $\sim 10^9$.  They also cross-correlate a small fraction of all antenna pairs, allowing the redundant baseline calibration software \cite{omnical,LOFARcal} to update the calibration coefficients in real time.
Finally, software running on regular computers combine all snapshots of sufficient quality into a 3D sky ball representing the sky brightness as a function of angle and frequency in Stokes (I,Q,U,V) \cite{FFTT2}, and subsequent software deals with foregrounds and measures power spectra and other cosmological observables.
\label{ArchitectureFig}
}
\end{figure}

Our architecture for a very large omniscope is shown in \fig{ArchitectureFig}. The fact that spatial correlations of antennas on a hierarchical grid can be performed by multidimensional FFTs is simply a mathematical theorem, and needs no experimental confirmation. On the other hand, it is crucial for our MITEoR prototype to demonstrate that  automatic precision calibration is possible using redundant baselines, since the calibration coefficients mentioned in the figure caption must be updated frequently to allow the FFTs to combine the signals from the different antennas without introducing errors. Because our digital hardware is powerful enough to allow it, our MITEoR prototype correlates all 128 input channels with one another, rather than just a small sample as mentioned in the caption.  This provides additional cross-checks that greatly aid technological development, where instrumentation may be particularly prone to systematics. In general, our mission is to empirically explore any challenges that are unique to a massively redundant interferometer array.  Once these are known, one can rather trivially reconfigure the cross-correlation hardware to perform spatial FFTs, thereby obtaining an omniscope with $N\log N$ correlator efficiency.

\subsection{The Analog System}

\begin{figure}
\centerline{\includegraphics[width=85mm]{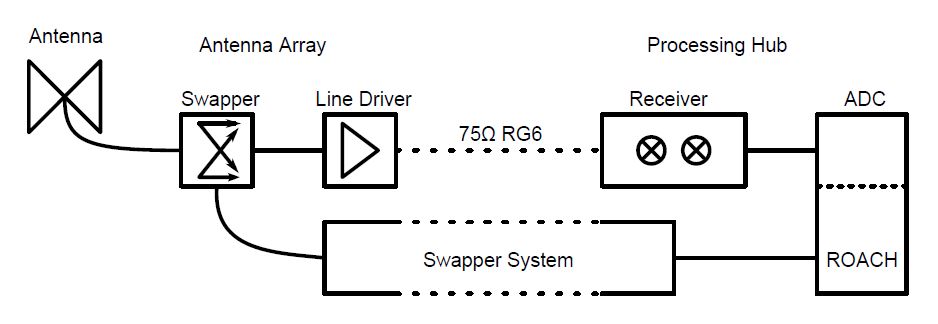}}
\caption{
System diagram of our analog system.\label{analog_schematic}
}
\end{figure}

MITEoR contains 64 dual-polarization antennas, giving 128 signal channels in total. The analog processing chain we designed consists of line-drivers, receivers and swappers, as shown in \fig{analog_schematic}. The signal picked up by the antennas is first amplified by two orders of magnitude by the low noise amplifiers (LNAs) built-in to the antennas. It is then phase switched in the swapper system, which greatly reduces cross-talk downstream. Following this, the signal is amplified again by about five orders of magnitude in our line-drivers before being sent over 50 meter RG6 cables to the receivers. The receivers perform IQ demodulation on a desired $50\,\textrm{MHz}$ band selected between $100\,\textrm{MHz}$ and $200\,\textrm{MHz}$, producing two channels with adjacent $25\,\textrm{MHz}$  bands, and sends the resulting signals into our digitization boards containing 256 analog-to-digital converters (ADCs) sampling at $50\,\textrm{MHz}$.  

When designing the components of this system, we chose to use commercially-purchased integrated circuits and filters when available, to allow us to focus on
system design and construction. In some cases (such as with the amplifiers) the cost of the IC is less than the cost of enough discrete transistors to implement even a rough approximation of the same functionality. Less expensive filters
could be made from discrete components, but the characteristics of purchased modules tend to be  better due to custom inductors and shielding. When we needed to produce our own boards as described below, our approach was to design, populate and test them in our laboratory, then have them affordably mass-produced for us by Burns Industries\footnote{ \url{http://www.burnsindustriesinc.com}}.

\subsubsection{Antennas}

The dual-polarization antennas used in MITEoR were originally developed for the Murchison Widefield Array \cite{MWA}, and consist of two ``bow-tie"-shaped arms as can be seen in \fig{ArrayFigure}.
The MWA antenna is inexpensive and the band of interest
is identical, so reuse of the design is appropriate. The MWA antenna was designed for a frequency range of $80$-$300\,\textrm{MHz}$, and
has a built-in low noise amplifier with $20\,\textrm{dB}$ of gain. The noise figure of the amplifier is $0.2\,\textrm{dB}$, and the $20\,\textrm{dB}$
of gain means that following gain stages do not contribute significantly to the noise figure.

\subsubsection{Swappers (phase switches)}

As for many other interferometers, crosstalk within the receivers, 
ADCs, and cabling significantly affect signal quality. 
We observe the cross-talk to depend strongly on the physical proximity of channel pairs, reaching as high as about $-40$dB between nearest neighbor  channels. Our swapper system is designed to cancel out crosstalk during the correlator's time averaging 
by selectively inverting analog signals using Walsh modulation. The channel from each antenna-polarization is inverted 50\% of the time according to a Walsh function, by an analog ZMAS-1 phase switch from Mini-Circuits located before the second amplification stage (line-driver),  
then appropriately re-inverted after digitization. This eliminates all crosstalk to first order. 
If cross-talk reduction were the only concern, the ideal position for the analog signal would be 
immediately after the antenna, in order to cancel as much crosstalk as possible. In practice, the swapper introduces a loss of about 3 dB, so we perform the modulation after the LNA to avoid adding noise (raising the system temperature). We have demonstrated that the swapper system attenuates crosstalk in our ADC by of order 40dB over the frequency band of interest, reducing it to being of order $- 80$dB for the most afflicted signal pairs.
 
 \vskip-1.1cm
 
 \begin{figure}[ht]
\vskip-1.5cm
\centerline{\includegraphics[width=110mm]{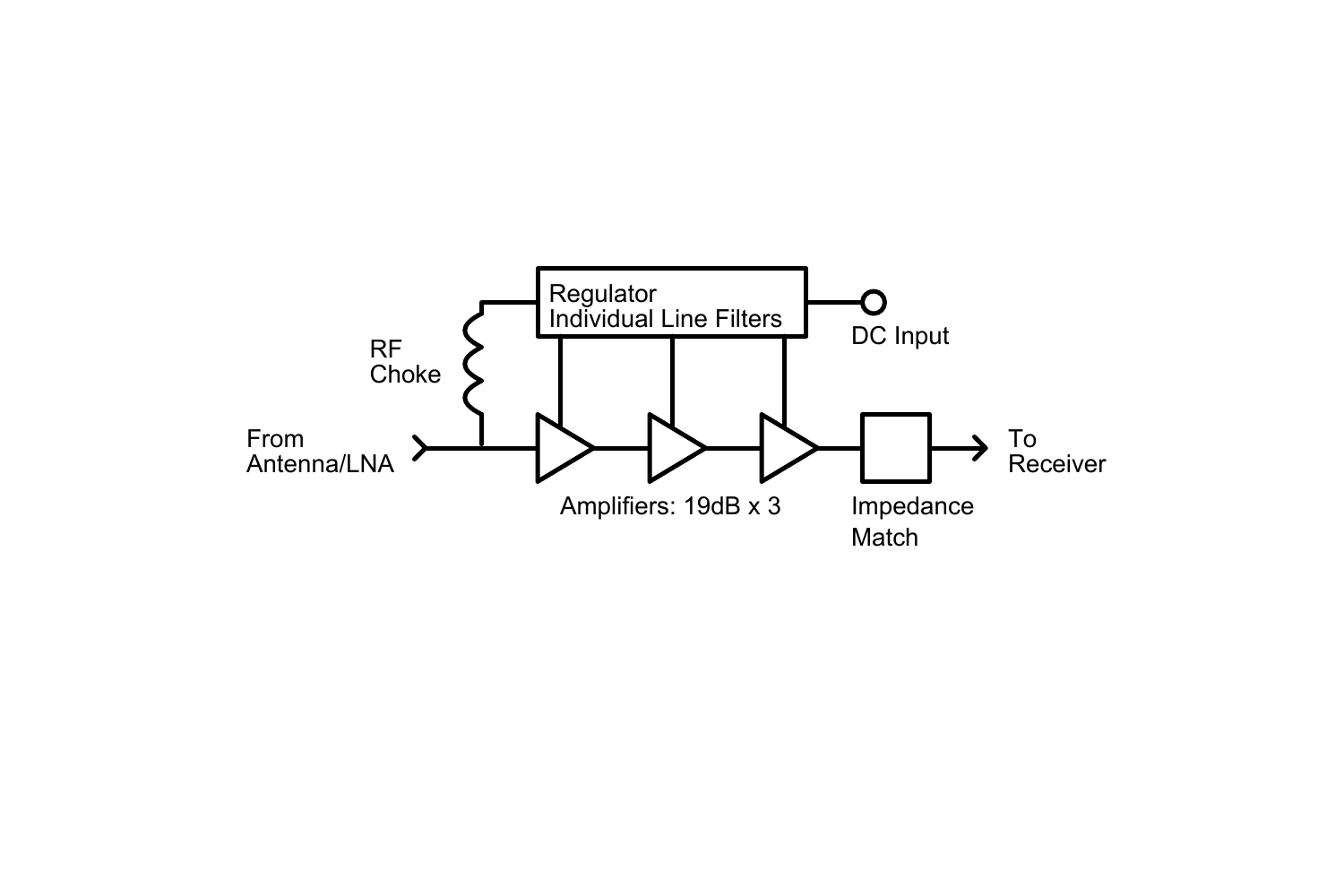}}
\vskip-2.5cm
\centerline{\includegraphics[width=80mm]{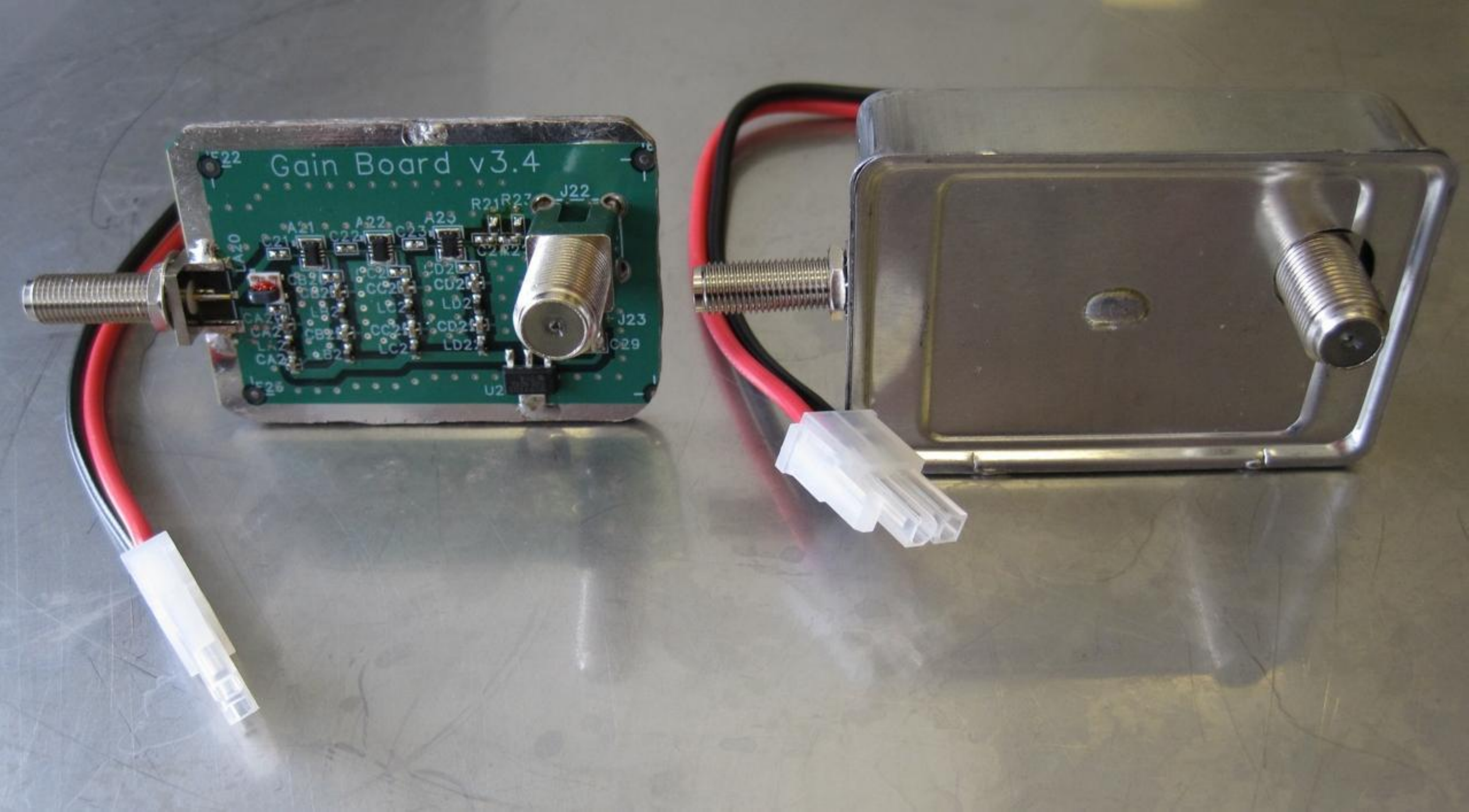}}
\caption{
The line driver that we have designed takes the signal in the 50 Ohm coaxial cable from the antenna LNA and amplifies it by 51 dB, in order to overpower noise picked up in the subsequent 75 Ohm coaxial cable and further processing steps up to 50 meters away.  It operates on 5V DC and also provides DC bias power to the antenna's LNA through the 50 Ohm cable.
\label{LineDriverFig}
}
\end{figure}

\subsubsection{Line-driver}
Line-drivers (\fig{LineDriverFig}) amplify a single antenna's signal from one of its two polarization channels while powering its LNA. Line-drivers only handle a single channel to reduce potential crosstalk from sharing a printed circuit board. They are placed near the antenna in order to reduce resistive
losses from powering the antenna at low voltage. Additional gain early in the analog chain helps the signal overpower
any noise picked up along the way to the processing hub, and maintains the low noise figure set up by the LNA. To further reduce potential radio-frequency interference (RFI), we chose to power the line-drivers with 58Ah 6V sealed lead acid rechargeable batteries during our final 64-antenna deployment, rather than 120VAC to 6VDC adapters (whose unwanted RF-emission may have caused occasional saturation problems during our earlier tests).

 \begin{figure}[ht]
\vskip-1.1cm
\centerline{\includegraphics[width=90mm]{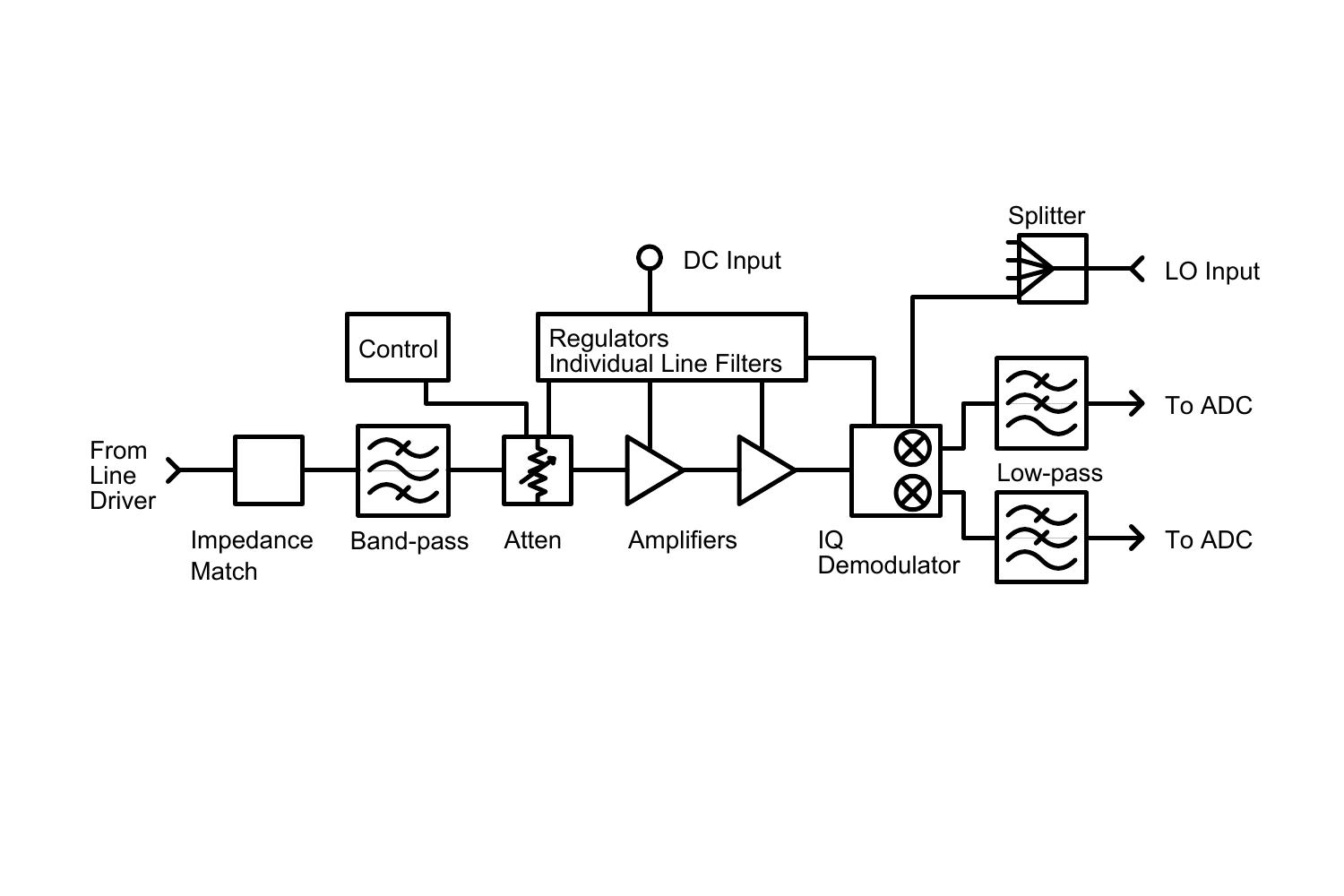}}
\vskip-1.9cm
\centerline{\includegraphics[width=80mm]{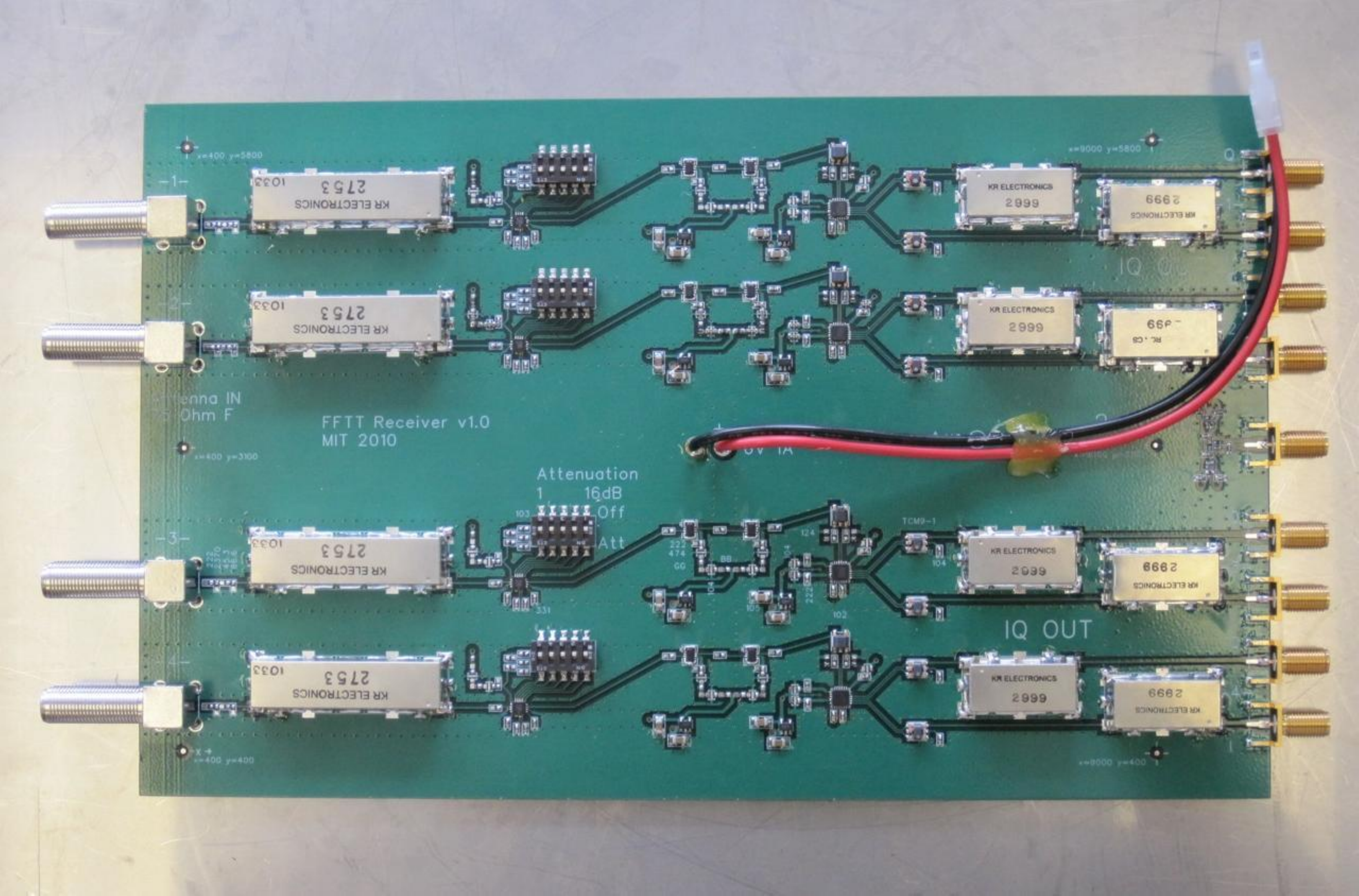}}
\caption{
The receiver boards that we have designed take the signals arriving from four line drivers, 
band-pass filter and amplify them, then frequency shifts them from the band of interest to a DC-centered signal suitable for ADC input. 
\label{ReceiverFig}
}
\end{figure}

\subsubsection{Receiver}

Our receivers take input from the line-drivers, bandpass filter the incoming signals, amplify their power level by 23 dB, and IQ-demodulate them. The resulting signals go directly to an ADC for digitization. Receivers are placed near the ADCs to which they are connected to reduce cabling for LO distribution and ADC connection. 
IQ demodulation is used, which doubles received bandwidth for a given ADC frequency at the cost of using two ADC channels, and has the advantage of requiring only a single local oscillator (LO) and low speed ADCs. The result is 40 MHz of usable bandwidth anywhere in the range 110-190 MHz, with a 2-3 MHz gap centered around the LO frequency.
The received boards have five pins allowing their signals to be attenuated by any amount between 0 dB and 31 dB (in steps of 1 dB) before the second amplification stage, to avoid saturation and attain signal levels optimal for digitization.
%

\subsection{The digital system}
\label{digi}
We designed MITEoR's digital system to be highly compact and portable, with the entire system occupying 3 shock-mounted equipment racks on wheels, each measuring about $5\,\textrm{ft}$ by $5\,\textrm{ft}$ by $3\,\textrm{ft}$. 
It takes in data from 256 ADC channels, Fourier
transforms the data into the frequency domain, reconstructs IQ demodulated channels back to 128 corresponding antenna channels, computes the cross-correlations of all pairs of the 128 antenna channels with 8 bit precision (rather than 4 bits seen in most correlators of similar specifications), and then time-averages these cross-correlations. The digital hardware is capable of processing an instantaneous bandwidth of $12.5\,\textrm{MHz}$ with frequency bin size $5\,\textrm{kHz}$, and the time interval for averaging and subsequent output to our data server is 
usually configured to be either 2.6 or 5.3 seconds.

\subsubsection{F-X design scheme}
We adopted the popular F-X scheme in MITEoR's digital system. We have 4 synchronized F-engines that take in data from 4 synchronized 64 channel ADC boards, which run at 14 bits and 50Ms/s. The F-engines perform the FFT and IQ reconstruction, and distribute the data onto 4 X-engines through 16 10GbE cables. The 4 asynchronous X-engines then perform full correlation on 4 different frequency bands on all 128 channels, and send the time averaged results to a computer for data storage.

To implement the computational steps of the MITEoR design, we used Field Programmable Gate Arrays (FPGAs). These devices can be programmed to function as dedicated pieces of computational hardware. Each F-engine and X-engine mentioned above is implemented by one FPGA. The FPGAs we use are seated on custom hardware boards developed by CASPER group at Berkeley \cite{4176933}. We also use the software tool flow developed by CASPER to design our digital system.


\subsubsection{CASPER Hardware}
The digital hardware platform of MITEoR has been designed by the
CASPER group at Berkeley. This group is dedicated to building open-source programmable hardware specifically for applications in astronomy. We currently use two of their newer
devices, the ROACH (Reconfigurable Open Architecture Computing Hardware) for our F-engines, and the ROACH 2 for our X-engines. They contain Xilinx Virtex FPGAs which are programmed
and controlled by a PowerPC processor running a Linux kernel. They also feature a
number of useful peripheral connections such as 10Gb Ethernet links and ports for
connecting high performance ADCs \cite{2008}. 

The main benefit of using CASPER hardware is that it eliminates a large part of
the design process for radio telescopes: namely, the amount of time it takes to design
and build custom hardware. Also, along with their hardware comes a large open-source
library of FPGA programming structures such as polyphase filter bank FIR filters and
fast Fourier transform blocks, which saves us from having to re-invent the
wheel, i.e., commonly used signal processing structures \cite{4840623}.

\section{MITEoR Observations}
\subsection{History}
We first built an $N=12$ array on the Kavli Institute rooftop at MIT and then tested an $N=16$ version at the Green Bank National Radio Observatory in August 2009, using MWA antennas, off-the-shelf USRP-receivers and software correlation. To improve bandwidth, cut per-element cost and make the architecture scalable, we then developed our own analog signal chain and switched to the open-source CASPER FPGA hardware platform. We finally completed our new hardware integration in September 2010, and performed a successful suite of test observations with an 8-antenna interferometer in The Forks, Maine (population 35, cell phone reception 0, a beautiful river valley 4.5 hours drive from MIT) which our online research suggests may be the most radio quiet area in all of the eastern United States.\footnote{The Forks has also been successfully used to test the Edges experiment \cite{EDGES}, and we found the RFI spectrum to be significantly cleaner than at the Green Bank National Radio Observatory at the very low (100-200 MHz) frequency range that is our focus.}

In 2012, we completed a major upgrade of the digital system to fully correlate $N=16$ dual polarization antennas, and had our 4th deployment in The Forks in May 2012. With the experience of this successful deployment, we further upgraded our digital system to accommodate $N=64$ dual polarization antennas, which led to our latest deployment in July 2013 described in the following subsection.

\begin{figure}[!ht]
\centerline{\includegraphics[width=85mm]{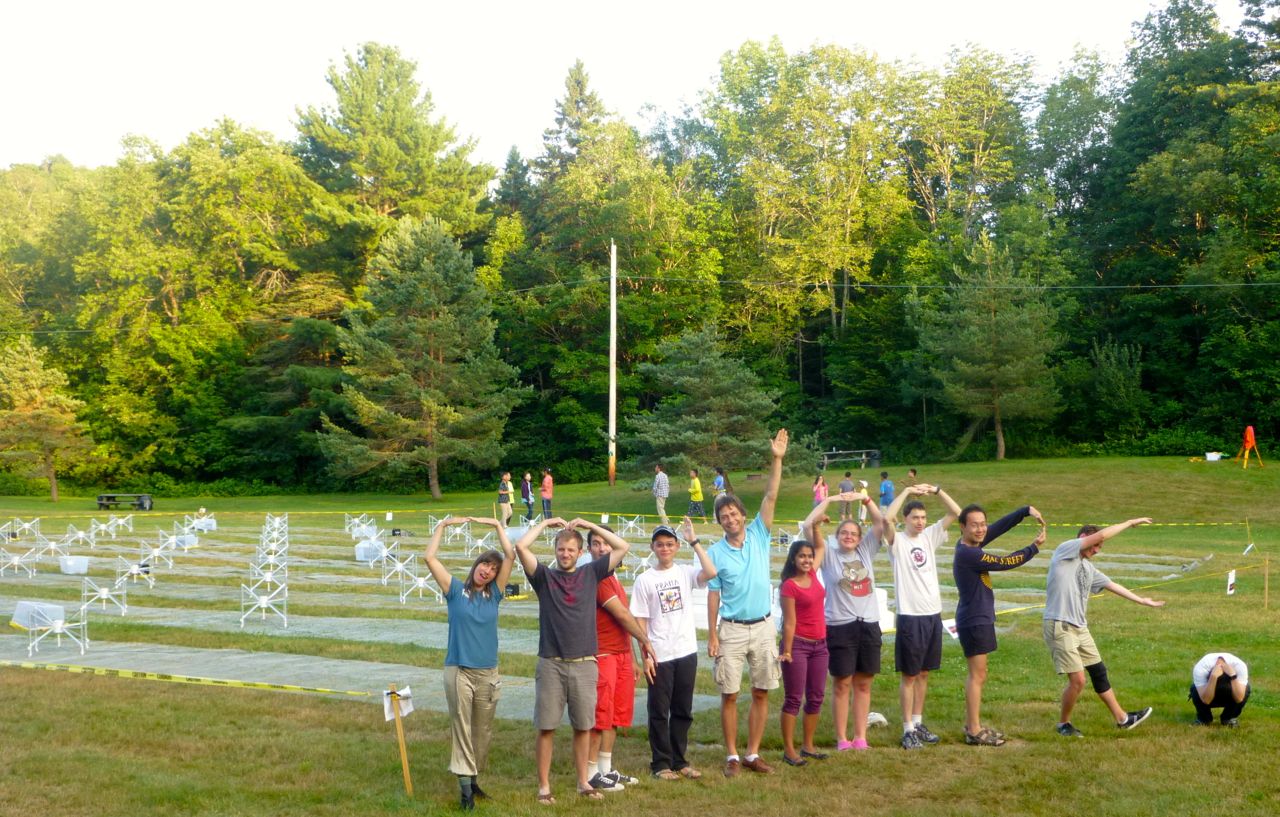}}
\caption{Part of the MITEoR array from our most recent deployment, with some of us attempting to spell ``Omniscope." in the foreground. The laser-ranging total station can be seen in the rear right.
\label{ArrayFigure}
}
\end{figure}

\subsection{Latest deployment}

On July 20th 2013, a crew of 11 of us drove from MIT to The Forks, Maine with the entire MITEoR experiment packed into a 17 foot U-haul truck. We started unpacking, assembling and deploying in the afternoon, and the whole 64 antenna system was up and running after two days. A skeleton crew of 3 members stayed on site for monitoring and maintenance for the following two weeks, during which we collected more than 300 hours of data. On August 4th, a demolition crew of 5 members disassembled and packed up MITEoR in 6 hours and concluded the successful deployment.

During the roughly 300 hours, we scanned through a frequency range of $123.5\,\textrm{MHz}$-$179.5\,\textrm{MHz}$, with at least 24 consecutive hours at each frequency. We used two different array layouts for most of the frequencies we covered. We began with the antennas arranged in a regular 8 by 8 grid, with 3 meter spacing between neighboring antennas, and reconfigured to an 8 by 8 regular grid with 1.5 meter spacing for a more compact layout (which provides better signal-to-noise ratio on the 21 cm signal). We measured and aligned the antenna positions using a laser-ranging total station. As can be seen in \fig{ArrayFigure}, strips of conducting ground screen were rolled out beneath the antennas to block thermal emission from the ground and to ensure that the primary beams are identical as possible for all antennas. 

The total volume of binary data collected was 3.9TB, and we are currently hard at work analyzing this data.

\section{Data calibration}

The key focus of MITEoR is precision calibration, because this is a prerequisite for successful use of the the spatial FFT algorithm. In this section we describe the calibration scheme that we have designed and implemented. The first step fixes the relative calibration between antennas, utilizing both per-baseline algorithms and our redundant-baseline calibration \cite{omnical}. The second step builds on the resulting calibration solution to constrain the absolute calibration, which involves both fixing the signal amplitudes to an astronomical flux scale and breaking a degeneracy in the orientation of the array.

\subsection{Relative calibration}
The goal of relative calibration is to calibrate out differences among antenna elements caused by non-identical analog components, such as variations in amplifier gains and cable lengths, which may be functions of time. We parametrize our calibration solution as a time- and frequency-dependent complex gain $g_i$ per antenna.  Calibrating our interferometer amounts to solving for the coefficients $g_i$ and undoing their effects on our data.

Suppose the $i$th antenna measures a signal $s_i$ at a given instant. This signal can be
written in terms of a complex gain factor $g_i$, the antenna's instrumental noise contribution $n_i$, and the true sky signal $x_i$ that would be measured in the limit of perfect gain and no noise:
\begin{equation}
s_i = g_i x_i + n_i.
\end{equation}
Each baseline measures the correlation between the two signals from the two participating antennas:
\begin{eqnarray}
\label{eq:eq1}
c_{ij} &\equiv& \langle s_i^* s_j \rangle \nonumber \\
&=&g_i^* g_j\langle x_i^* x_j\rangle +\langle n_i^* n_j\rangle + g_i^*\langle x_i^* n_j\rangle + g_j\langle n_i^* x_j\rangle \nonumber \\ 
&=&g_i^* g_j\langle x_i^* x_j\rangle + n_{ij}^{res}\equiv g_i^* g_jy_{i-j} + n_{ij}^{res}
\end{eqnarray}

where we have denoted the true correlation $\langle x_i^* x_j\rangle$ by $y_{i-j}$, and the angled brackets $\langle \dots \rangle$ denote time averages. 

As a first step in our scheme for the relative calibration, we make crude estimates of the quantity $g_i^* g_j$ by assuming knowledge of $y_{i-j}$. We do so by simulating $y_{i-j}$ using the Global Sky Model presented in \cite{GSM}. Since the Global Sky Model is not a perfect model of the real sky, the $g_i$s we obtain with this method may not represent the best calibration solution. Nevertheless, they provide a sufficiently good starting point for our next step.

The second step of our relative calibration is known as redundant baseline calibration. In a maximally redundant array such as the MITEoR, the number of unique baselines is much less than the total number of baselines. Therefore, we can treat all the $y_{i-j}$s as unknowns in addition to the $g_i$s, and the system of equations \eqref{eq:eq1} is still overdetermined, enabling us to find best fits for the  $y_{i-j}$s and the $g_i$s. Our current redundant calibration algorithm (the ``log calibration" algorithm of \cite{omnical}) requires a close first guess to the $g_i$s to break phase-wrapping degeneracies, hence the need to perform the aforementioned first step prior to this.

\begin{figure}[!ht]
\centerline{\includegraphics[width=80mm]{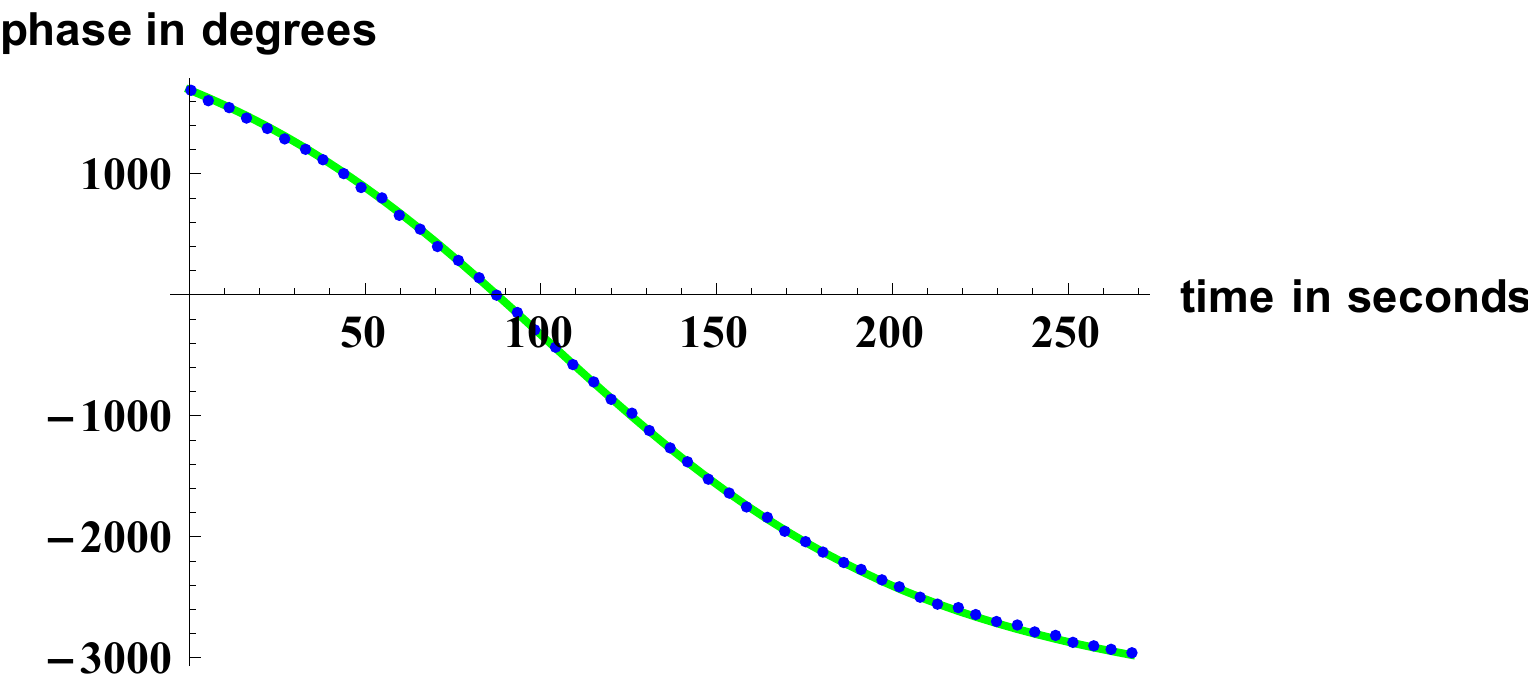}}
\caption{
Calibrated phase measurements of a single ORBCOMM satellite as a function of time observed by a pair of antennas. In black are the calibrated measurements, and in green is a theoretical prediction.
\label{sat4}
}
\end{figure}

\subsection{Absolute calibration}
The absolute calibration of the instrument involves two separate aspects. One is to find the overall gain of all antennas over frequency and time, and the other to calibrate the overall orientation of the array.  The former is equivalent to fixing our signal amplitudes to an astronomical flux scale, which can be dealt with by comparing our data to published astronomical catalogs (see \cite{dannyCal} for example).  As this is not a problem unique to calibrating massively redundant arrays, we defer a description of this process to future papers.  Here, we focus on the second problem, which occurs whenever one employs calibration schemes that rely solely on baseline redundancy.  Since such algorithms use only the internal consistency of an interferometric system to solve for calibration solutions, they are invariant under a tilting of the entire array.  External data must therefore be used to fix the orientation of the array.

To solve this problem for MITEoR, we make use of the ORBCOMM satellites, which emit strongly (typically 100 times stronger than the brightest celestial sources) at a few specific frequencies in our frequency band.  Because of this brightness, plus the fact that their quickly-varying sky positions are available in publicly-accessible archives, ORBCOMM satellites are perfect calibrators for  constraining the tilt of our array.  With the 8-bit design of the correlator described in Section \ref{digi}, MITEoR is able to calibrate off ORBCOMM satellites in an \emph{in situ} fashion,  making calibration measurements and astronomical observations simultaneously.

In \fig{sat4} we show the calibrated phase measurement of a single ORBCOMM satellite (which depends on its angular position) as a function of time, after applying our redundant-baseline calibration algorithm.  A theoretical prediction for the expected signal is plotted in light green, and one sees that the post-calibration data is in excellent agreement with theory.  Using this satellite signal, we can determine our array orientation in all three directions to much better than one degree, which is more than sufficient for our needs given the interferometer's angular resolution.

\section{Summary and outlook}

We have described the MITEoR experiment, a pathfinder ``omniscope'' radio interferometer with 64 dual-polarization antennas in a highly redundant configuration. Although the data analysis from our summer 2013 deployment has only just begun, it is clear that the massively redundant baseline calibration is working exquisitely. This bodes well for future attempts to perform such calibration in real-time instead in post-processing, thus clearing the way for FFT correlation that will cut the cost of correlating $N$ antennas from scaling as $N^2$ to scaling as $N\log N$.  It also suggests that the extreme calibration precision required to reap the full potential of 21 cm cosmology is within reach. 

The omniscope architecture that MITEoR successfully demonstrates is now being incorporated into the much more ambitious Hydrogen Epoch of Reionization Array (HERA), a broad-based collaboration among US radio astronomers from the PAPER, MWA and MITEoR experiments. HERA plans to deploy about 600 14-meter dishes in a close-packed hexagonal array in South Africa, giving a collecting area around 0.1 square kilometers, virtually guaranteeing not only a solid detection of the elusive cosmological 21 signal but also interesting new clues about our cosmos.

{\bf Acknowledgments:}

MITEoR was supported by NSF grants AST-0908848 and AST-1105835, the MIT Kavli Instrumentation Fund, the MIT undergraduate research opportunity (UROP) program, FPGA donations from XILINX, and by generous donations from Jonathan Rothberg and an anonymous donor. We wish to thank Jacqueline Hewitt and Aaron Parsons for helpful comments and suggestions, Dan Werthimer and his CASPER group for developing and sharing their hardware and teaching us how to use it, Jean Papagianopoulos, Thea Paneth, Jonathan Kocz, Steve Burns and his family for invaluable help, Meia Chita-Tegmark, Philip and Alexander Tegmark and Sherry Sun for deployment assistance,
and Joe Christopher and his fellow denizens of The Forks for their awesome hospitality and support.

\bibliography{omniproceedings}
\end{document}